\documentclass[pra,onecolumn,showpacs,showkeys,preprintnumbers,amsmath,amssymb,floatfix]{revtex4}

  \usepackage[dvips]{graphicx}
\usepackage{dcolumn}
\usepackage{bm}
\usepackage{axodraw}

\begin{document}


%
%
\title{X-ray energies of circular transitions and electrons screening
  in kaonic atoms}

%
%
\author{J.\ P.\ Santos}
\email{jps@cii.fc.ul.pt}
\affiliation{Departamento de F{\'\i}sica, Faculdade de Ci{\^e}ncias e Tecnologia, \\
  Universidade Nova de Lisboa, Monte de Caparica, 2825-114 Caparica, Portugal, \\
  and Centro de F{\'\i}sica At{\'o}mica da Universidade de Lisboa, \\
  Av. Prof. Gama Pinto 2, 1649-003 Lisboa, Portugal}

\author{F.\ Parente}%
\affiliation{Departamento F{\'\i}sica da Universidade de Lisboa \\
   and Centro de F{\'\i}sica At{\'o}mica da Universidade de Lisboa, \\
   Av. Prof. Gama Pinto 2, 1649-003 Lisboa, Portugal}

\author{S.\ Boucard}
\affiliation{Laboratoire Kastler Brossel,
  \'Ecole Normale Sup\' erieure et Universit\' e P. et M. Curie\\
  Case 74, 4, place Jussieu, 75252 Paris CEDEX 05, France}
%

\author{P.\ Indelicato} 
\email{paul.indelicato@spectro.jussieu.fr}
\affiliation{Laboratoire Kastler Brossel,
  \'Ecole Normale Sup\' erieure et Universit\' e P. et M. Curie\\
  Case 74, 4, place Jussieu, 75252 Paris CEDEX 05, France}
\author{J.\ P.\ Desclaux} 
\affiliation{15 Chemin du Billery
 F-38360 Sassenage,  France}

\date{\today}

%
%

\begin{abstract}
  The QED contribution to the energies of the circular $(n,\ell
  =n-1)$, $2 \le n \le 19$ transitions have been calculated for
  several kaonic atoms throughout the periodic table, using the
  current world average kaon mass.  Calculations were done in the
  framework of the Klein-Gordon equation, with finite nuclear size, finite particle size, and
  all-order Uelhing vacuum polarization corrections, as well as
  K\"all\'en and Sabry and Wichmann and Kroll corrections.  These energy
  level values are compared with other computed values. The circular
  transition energies are compared with available measured and
  theoretical transition energy. Electron screening is evaluated using
  a Dirac-Fock model for the electronic part of the wave function. The
  effect of electronic wavefunction correlation is evaluated for the
  first time.
\end{abstract}
%
%

\pacs{36.10.-k, 36.10.Gv, 32.30.Rj}
%


\keywords{kaonic atoms; electron screening}
\maketitle

%
%
\section{\label{sec_introduction}Introduction}

An exotic atom is formed when a particle, with a negative charge and
long-enough lifetime, slows down and stops in matter. It can then
displace an atomic electron, and become bound in a high principal
quantum number atomic orbital around the nucleus. The principal
quantum number of this highly excited state is of the order of
$n=\sqrt{m/m_{e}}$, where $m$ and $m_{e}$ are the masses of the
particle and of the electron, respectively \cite{700}. The higher the
overlap between the wave functions of the electron and the particle,
the more probable is the formation of an exotic atom \cite{700}.

The exotic atoms formed in this way are named after the particle
forming them. It the particle is a the negative kaon $K^{-}$, a meson
with a spin-0 and a lifetime of 1.237$\times 10^{-8}$ s, a kaonic atom
is thus created.

Because the particle mass, and thus transition energies are so much
higher that the electron's (a kaon is $\approx$964 times heavier than
an electron), the de-excitation of the exotic atom will start via
Auger processes, in a process equivalent to internal conversion for
$\gamma$-rays, while the level spacing is small and there are
electrons to be ejected, and then via radiative ($E1$) transitions,
producing characteristic X-rays while cascading down its own sequence
of atomic levels until some state of low principal quantum number. One
thus can end with a completely striped atom, provided the mass of the
exotic particle is large and the atomic number of the atom not too
high.

The initial population of the atomic states is related to the
available density of states, so for any given principal quantum number
$n$ the higher orbital momenta are favored to some extent because of
their larger multiplicity. As the Auger transitions do not change the
shape of the angular momentum distribution, the particle quickly
reaches the ($\ell =n-1$) orbits \cite{700}. Once the radiative ($E1$)
transitions begin to dominate, we have the selection rule $\Delta \ell
=\pm 1$, often with many possible values of $\Delta n$ being
important. Under such a scheme, the kaons in low angular momentum
orbits will rapidly reach orbitals with a sizable overlap with the
nucleus and be captured.  Soon mostly the circular orbitals $(n,\ell
=n-1)$ from which only transitions to other circular orbits can occur,
($n$, $\ell =n-1$) $\rightarrow $($n-1$, $\ell =n-2$), will be
populated. The so-called parallel transition ($n$, $\ell =n-2$)
$\rightarrow $($n-1$, $\ell =n-3$) is much weaker.

Finally the particle in a state of low angular momentum will be
absorbed by the nucleus through the kaon-nucleus strong interaction.
This strong interaction causes a shifting of the energy of the lowest
atomic level from its purely electro-magnetic value while the
absorption reduces the lifetime of the state and so X-ray transitions
to this final atomic level are broadened.

Therefore, following the stopping of the kaon in matter, well-defined
states of a kaonic atom are established and the effects of the
kaon-nucleus strong interaction can be studied. The overlap of the
atomic orbitals with the nucleus covers a wide range of nuclear
densities thus creating a unique source of information on the density
dependence of the hadronic interaction.

Here we are concerned with levels in which the effect of the strong
interaction is negligible, to study the atomic structure. Our
objective is to provide highly accurate values of high-angular
momentum circular transition, which can be useful for experiment in
which internal calibration lines, free from strong interaction shifts
and broadening are needed, as has been done in the case of pionic and
antiprotonic atoms \cite{779,729,778,870}.  Our second objective is to
study the electron screening, i.e., the change in the energy of the
kaon due to the presence of a few remaining electrons, in a
relativistic framework.

%
%
%
%
In general, although the kaonic atom energy is dominated by the
Coulomb interaction between the hadron and the nucleus, one must take
into account the strong-interaction between the kaon and the nucleus
when the kaon and nuclear wave function overlap.  As our aim is to
provide highly accurate QED-only values, that can be used to extract
experimental strong-interaction shifts from energy measurements. We
refer the interested reders to the literature, e.g.,
Refs.~\cite{678,677,759}.

Exotic X-ray transitions have been intensively studied for decades as
they can provide the most precise and relevant physical information
for various fields of interest. This study is pursued at all major,
intermediate energy accelerators where mesons are produced: at the AGS
of Brookhaven National Laboratory (USA), at JINR (Dubna, Russia), at
LAMPF, the Los Alamos Meson Physics Facility (Los Alamos, USA), at
LEAR, the Low Energy Antiproton Ring of CERN (Geneva, Switzerland), at
the Meson Science Laboratory of the University of Tokyo (at KEK,
Tsukuba, Japan), at the Paul Scherrer Institute (Villigen,
Switzerland), at the Rutherford Appleton Laboratory (Chilton,
England), at the Saint-Petersburg Nuclear Physics Institute (Gatchina,
Russia), and at TRIUMF (Vancouver, Canada) \cite{700}.

It is now planed, in the context of the DEAR experiment (strong
interaction shift and width in kaonic hydrogen) on DA$\Phi$NE in Frascati
\cite{687,697} to measure some other transitions of
kaonic atoms, such as the kaonic nitrogen, aluminum, titanium, neon,
and silver. Similar work is under way at KEK.

The measurements concerning the exotic atoms permit the extraction of
precise information about the orbiting particle \cite{754}, such as
charge/mass ratio, and magnetic moment, and the interaction of such
particles with nuclei. In addition, properties of nuclei \cite{757},
such as nuclear size, nuclear polarization, and neutron halo effects
in heavy nuclei \cite{756}, have been studied. Furthermore, the
mechanism of atomic capture of such heavy charged particles and atomic
effects such as Stark mixing \cite{758} and trapping \cite{755} have
been studied extensively.

Furthermore, as the X-ray transition energies are proportional to the
reduced mass of the system, studies in the intermediate region of the
atomic cascade were used to measure the masses of certain negative
particles. In order to minimize the strong interactions for the mass
determination it is considered only transitions between circular
orbits ($n,n-1$)$\rightarrow $($n-1,n-2$) far from the nucleus
\cite{707}.

In this paper, we calculate the orbital binding energies of the kaonic
atoms for $1 \leq Z \leq 92$ and $0 \leq \ell \leq 12$ for circular
states and $1 \leq \ell \leq 11$ for parallel near circular states by
the resolution of the Klein-Gordon equation (KGE) including QED
corrections.

The paper is organized as follows. The principle of the calculation is
outlined in Sec. \ref{sec:calculation}. The results obtained in this
work are given in Sec. \ref{sec:res_dis}. Finally we give our
conclusions in Sec. \ref{sec:conclusion}.

%
%
%
\section{\label{sec:calculation}Calculation of the energy levels}

%
%
\subsection{Principle of the calculation}
\label{subsec:principle}

Because of the much larger mass of the particle, its orbits are much
closer to the central nucleus than those of the electrons. In
addition, since there is only one heavy particle, the Pauli principle
does not play a role and the whole range of classical atomic orbits
are available. As a result, to first order, the outer electrons can be
ignored and the exotic atom has many properties similar to those of
the simple, one-electron hydrogen atom \cite {690}. Yet there are cases where the interaction
between the electron shell and the exotic particle must be considered. 
Over the year the MCDF code of Desclaux and Indelicato \cite{32,282} was modified so that 
it could accommodate a wavefunction that is the product of a Slater determinant for the electron, by 
the wavefunction of an exotic particle. In the case of spin 1/2 fermions, this can be done with the full Breit interaction. In the case of spin 0 bosons, this is restricted to
the Coulomb interaction only. With that code it is then possible to investigates the effects of changes in the electronic wavefunction, e.g., due to correlation on the
kaonic transitions. It is also possible to take into account specific properties of the exotic particle like its charge distribution radius. The different 
contributions to the final energy are described in more details in this section.


%
%
%
\subsection{\label{subsec:KGE} Numerical solution of the Klein-Gordon equation}

Since kaons are spin-0 bosons, they obey the Klein Gordon equation, which in the
absence of strong interaction may be written, in atomic units, as
%
\begin{equation}
\left[\alpha^2\left(E-V_{c}(r) 
         \right)^{2}+{\vec 
         \nabla}^{2}-\mu^{2}c^{2}\right]\psi\left(\mathbf{r}\right)=0,
\label{eq:KGE}
\end{equation}
where $\mu $ is the kaon reduced mass, $E$ is the Kaon total energy,
$V_{c}$ is the sum of the Coulomb potential, describing the
interaction between the kaon and the finite charge distribution of the
nucleus, of the Uehling vacuum-polarization potential (of order
$\alpha \left( \alpha Z\right)$) \cite{ueh35} and of the potential due to the
electrons.  Units of $\hbar =c=1$ are used.  For a spherically
symmetric potential $V_{c}$, the bound state solutions of the KG
equation (\ref{eq:KGE}) are of the usual form $\psi _{n\ell m}\left(
  \mathbf{r}\right) =Y_{\ell m}\left( \theta ,\phi \right)\left(
  p_{n\ell }\left( r\right) /r\right) $. Compared to the numerical
solution of the Dirac-Fock equation \cite{282}, here we must take care
of the fact that the equation is quadratic in energy. The radial
differential equation deduced from (\ref{eq:KGE}) is rewritten as a
set of two first-order equations
\begin{eqnarray}
  \label{eq:kg1st}
  \frac{d}{dr} p & = &q \nonumber \\
  \frac{d}{dr} q &= &\left[\mu c^2 + \frac{\ell (\ell + 1)}{r^2} -\alpha^2 \left(V_{c} -E\right)^2\right]p,
\end{eqnarray}
where $p=p(r)$ is the radial KG wave function. Following
Ref.~\cite{282} the equation is solved by a shooting method, using a
predictor-corrector method for the outward integration up to a point
$r_m$ which represents the classical turning point in the potential
$V_c$. The inward integration uses finite differences and the tail
correction and provide a continuous $p$ as well as a practical way to
fix how far out one must start the integration to solve Eq.~
(\ref{eq:kg1st}) within a given accuracy. The eigenvalue $E$ is found
by requesting that $q$ is continuous at $r_m$. If we suppose that $q$
is not continuous, then an improved energy $E$ is obtained by a
variation of $p$ and $q$, such that
\begin{equation}
  \label{eq:var1}
  (q+\delta q)_{r_m^+} = (q+\delta q)_{r_m^-}.
\end{equation}
To find the corresponding $\delta E$ we replace $p$, $q$ and $E$ by
$p+\delta p$, $q+\delta q$ and $E+\delta E$ in Eq. (\ref{eq:kg1st}).
Keeping only first order terms we get
\begin{eqnarray}
  \label{eq:kg1stexp}
  \frac{d}{dr} p+ \frac{d}{dr} \delta p & = &q +\delta q \\
  \label{eq:kg1stexp2}
  \frac{d}{dr}\delta q &= &\left[\mu c^2 + \frac{\ell (\ell + 1)}{r^2} -\alpha^2 \left(V_{c} -E\right)^2\right]\delta p\nonumber \\
 && + 2\alpha^2 (V-E)\delta E p.
\end{eqnarray}
Multiplying Eq. (\ref{eq:kg1stexp}) by $q$ and
Eq.~(\ref{eq:kg1stexp2}) by $p$, subtracting the two equations, and
using the original differential equation (\ref{eq:kg1st}) whenever
possible we finally get
\begin{equation}
  \label{eq:vardif}
   \frac{d}{dr}(p\delta q - q \delta p ) = 2 \alpha^2 \left(V_{c} -E\right)  p^2\delta E.
\end{equation}
Combining Eqs. (\ref{eq:var1}), (\ref{eq:vardif}), integrating, using
the fact that $p$ is continuous everywhere, and neglecting
higher-order corrections, we finally get
\begin{equation}
  \label{eq:deltae}
  \delta E = \frac{p(r_m) \left[q(r_m^+) - q(r_m^-)
  \right]}{2\alpha^2\int_{0}^{\infty} \left(V_{c} -E\right)  p^2 dr}. 
\end{equation}
In the case of the Dirac equation one would get 
\begin{equation}
  \label{eq:deltaedir}
  \delta E = \frac{p(r_m) \left[q(r_m^+) - q(r_m^-)
  \right]}{\alpha\left[\int_{0}^{r_m^-} ( p^2 + q^2) dr
  +\int_{r_m^+}^{\infty} ( p^2 + q^2) dr\right]}, 
\end{equation}
where $p$ and $q$ are the large and small components, and the integral
in the denominator remain split because $q$ is not continuous, but is
converging toward 1, as it is the norm of the wave function.  In both
cases one can use Eq.~(\ref{eq:deltae}) or (\ref{eq:deltaedir}) to
obtain high-accuracy energy and wave function by an iterative
procedure, checking the number of node to insure convergence toward
the right eigenvalue.

%
%
%
\subsection{\label{subsec:nuclear}Nuclear structure}

For heavy elements, a change between a point-like and an extended
nuclear charge distribution strongly modifies the wave function near
the origin.  One nuclear contribution is easily calculated by using a
finite charge distribution in the differential equations from which
the wave function are deduced.  For atomic number larger than 45 we
use a Fermi distribution with a thickness parameter $t=2.3$~fm and a
uniform spherical distribution otherwise. The most abundant
naturally-occurring isotope was used.

In Table \ref{tab:nuclear_parameters} we list the nuclear
parameters used in the presented calculations in atomic units.

\begin{table*}                      
{\footnotesize                      
\begin{center}                      
\caption{Nuclear parameters used in this work in atomic units                      
\protect\label{tab:nuclear_parameters}}                              
\begin{tabular}{rccc}                              
\hline                              
\hline                              
\\                              
Z  &   Nuclear Radius    &  Mean Square Radius     &  Reduced Mass    \\        
\\                              
\hline                              
\\                              
 1 & $ 2.102958 \times 10^{ -5 } $ & $ 1.628944 \times 10^{ -5 } $ & $ 633.030  $  \\        
 2 & $ 4.081494 \times 10^{ -5 } $ & $ 3.161512 \times 10^{ -5 } $ & $ 853.110  $  \\          
 3 & $ 5.838025 \times 10^{ -5 } $ & $ 4.522115 \times 10^{ -5 } $ & $ 898.234  $  \\          
 4 & $ 6.145418 \times 10^{ -5 } $ & $ 4.760220 \times 10^{ -5 } $ & $ 912.431  $  \\          
 6 & $ 5.989282 \times 10^{ -5 } $ & $ 4.639278 \times 10^{ -5 } $ & $ 925.227  $  \\          
 13 & $ 7.418903 \times 10^{ -5 } $ & $ 5.746657 \times 10^{ -5 } $ & $ 947.486  $  \\          
 14 & $ 7.579918 \times 10^{ -5 } $ & $ 5.871379 \times 10^{ -5 } $ & $ 948.135  $  \\          
 17 & $ 8.136153 \times 10^{ -5 } $ & $ 6.302237 \times 10^{ -5 } $ & $ 951.674  $  \\          
 19 & $ 8.384263 \times 10^{ -5 } $ & $ 6.494422 \times 10^{ -5 } $ & $ 953.134  $  \\          
 20 & $ 8.496485 \times 10^{ -5 } $ & $ 6.581349 \times 10^{ -5 } $ & $ 953.453  $  \\          
 22 & $ 8.780214 \times 10^{ -5 } $ & $ 6.801124 \times 10^{ -5 } $ & $ 955.537  $  \\          
 28 & $ 9.241059 \times 10^{ -5 } $ & $ 7.158094 \times 10^{ -5 } $ & $ 957.342  $  \\          
 29 & $ 9.509662 \times 10^{ -5 } $ & $ 7.366153 \times 10^{ -5 } $ & $ 958.031  $  \\          
 42 & $ 1.073460 \times 10^{ -4 } $ & $ 8.314984 \times 10^{ -5 } $ & $ 960.899  $  \\          
 45 & $ 1.098320 \times 10^{ -4 } $ & $ 8.507547 \times 10^{ -5 } $ & $ 961.150  $  \\          
 60 & $ 1.101005 \times 10^{ -4 } $ & $ 9.287059 \times 10^{ -5 } $ & $ 962.506  $  \\          
 74 & $ 1.221443 \times 10^{ -4 } $ & $ 1.015048 \times 10^{ -4 } $ & $ 963.325  $  \\          
 82 & $ 1.256895 \times 10^{ -4 } $ & $ 1.040691 \times 10^{ -4 } $ & $ 963.645  $  \\      
 92 & $ 1.348635 \times 10^{ -4 } $ & $ 1.107455 \times 10^{ -4 } $ & $ 963.955  $  \\      
\\                            
\hline                            
\hline                            
\end{tabular}                            
\end{center}                            
}                            
\end{table*}                            
%
%
\subsection{\label{subsec:qed}QED effects}

%
%
%
\subsubsection{Self-consistent Vacuum Polarization}
\label{subsubsec:vpsc}

A complete evaluation of radiative corrections in kaonic atoms is
beyond the scope of the present work.  However the effects of the
vacuum polarization in the Uehling approximation, which comes from
changes in the bound-kaon wave function, can be relatively easily
implemented in the framework of the resolution of the KGE using a
self-consistent method.  

\begin{figure}[htb]
\centering
\setlength{\unitlength}{0.27 mm}
\begin{picture}(260,100)(0,0)
\SetScale{.8}
%
\SetOffset(-30,35)
  \Line(0,0)(0,50)
  \Line(3,0)(3,50)
  \Vertex(1.5,25){2}
  \Vertex(28.5,25){2}
  \Photon(3,25)(27,25){3}{3}
  \CArc(38.5,25)(10,0,360)
  \Vertex(48.5,25){2}
  \DashLine(48.5,25)(68.5,25){3}
  \Text(74.5,26)[]{$\times$}
  \Text(90,26)[c]{+}
  \Text(40,-25)[c]{A1}
%
%
\SetOffset(75,35)
  \Line(0,-5)(0,55)
  \Line(3,-5)(3,55)
  \Text(90,26)[c]{+}
  \Text(40,-25)[c]{A2}
\SetOffset(75,50)
  \Vertex(1.5,25){2}
  \Vertex(28.5,25){2}
  \Photon(3,25)(27,25){3}{3}
  \CArc(38.5,25)(10,0,360)
  \Vertex(48.5,25){2}
  \DashLine(48.5,25)(68.5,25){3}
  \Text(74.5,26)[]{$\times$}
\SetOffset(75,20)
  \Vertex(1.5,25){2}
  \Vertex(28.5,25){2}
  \Photon(3,25)(27,25){3}{3}
  \CArc(38.5,25)(10,0,360)
  \Vertex(48.5,25){2}
  \DashLine(48.5,25)(68.5,25){3}
  \Text(74.5,26)[]{$\times$}
%
%
\SetOffset(180,20)
  \Line(0,0)(0,80)
  \Line(3,0)(3,80)
\SetOffset(180,35)
  \Vertex(1.5,25){2}
  \Vertex(28.5,25){2}
  \Photon(3,25)(27,25){3}{3}
  \CArc(38.5,25)(10,0,360)
  \Vertex(48.5,25){2}
  \DashLine(48.5,25)(68.5,25){3}
  \Text(74.5,26)[]{$\times$}
\SetOffset(180,5)
  \Vertex(1.5,25){2}
  \Vertex(28.5,25){2}
  \Photon(3,25)(27,25){3}{3}
  \CArc(38.5,25)(10,0,360)
  \Vertex(48.5,25){2}
  \DashLine(48.5,25)(68.5,25){3}
  \Text(74.5,26)[]{$\times$}
\SetOffset(180,65)
  \Vertex(1.5,25){2}
  \Vertex(28.5,25){2}
  \Photon(3,25)(27,25){3}{3}
  \CArc(38.5,25)(10,0,360)
  \Vertex(48.5,25){2}
  \DashLine(48.5,25)(68.5,25){3}
  \Text(74.5,26)[]{$\times$}
\SetOffset(180,60)
  \Text(85,1)[l]{+ $\cdots$}
  \Text(40,-50)[c]{A3}
\end{picture}
        \caption[]{Feynman diagrams obtained when the Uehling term
          is added to the nuclear potential; A1, A2 and A3 are,
          respectively, contributions of order of $\alpha(\alpha Z)$,
          $[\alpha(\alpha Z)]^2$ and $[\alpha(\alpha Z)]^3$. The
          dashed lines starting with a $\times$ represent the
          interaction with the nucleus, the double line a bound
          kaon wave function or propagator and the wavy line a
          retarded photon propagator.}
\label{fig_vpsc11}
\end{figure}
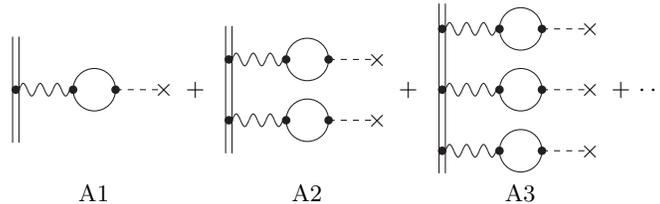

In practice one only need to add the Uehling potential to the
nuclear Coulomb potential, to get the contribution of the vacuum
polarization to the wave function to all orders, which is equivalent
to evaluate the contribution of all diagrams with one or several
vacuum polarization loop of the kind displayed on
Fig.~\ref{fig_vpsc11}. For the exact signification of these diagrams
see, {\em e.g.}, \cite{772,773,420}.

This happens because the used self-consistent method is based on a
direct numerical solution of the wave function differential equation.
Many precautions must be taken however to obtain this result as the
vacuum polarization potential is singular close to the origin, even
when using finite nuclei. The method used here is described in detail
in Ref.~\cite{bai2000}, and is based on \cite{kla77} and numerical coefficients found in \cite{277}.

%
%
%

\begin{figure}[htb]
\centering
\setlength{\unitlength}{0.27 mm}
\begin{picture}(260,50)(0,0)
\SetScale{.8}
%
\SetOffset(100,0)
  \Line(0,0)(0,50)
  \Line(3,0)(3,50)
  \Vertex(1.5,25){2}
  \Vertex(28.5,25){2}
  \Photon(3,25)(27,25){3}{3}
  \CArc(38.5,25)(10,0,360)
  \Vertex(48.5,25){2}
  \DashLine(48.5,25)(67,25){3}
  \Text(73,26.5)[]{$\times$}
  \Vertex(38.5,15){2}
  \DashLine(38.5,15)(67,15){3}
  \Text(73,16)[]{$\times$}
  \Vertex(38.5,35){2}
  \DashLine(38.5,35)(67,35){3}
  \Text(73,37)[]{$\times$}

%
\end{picture}
        \caption[]{Feynman diagrams for the Wichmann-Kroll potential
          of the order $\alpha(\alpha Z)^3$. The dashed lines starting
          with a $\times$ represent the interaction with the nucleus,
          the double line a bound kaon wave function or propagator
          and the wavy line a retarded photon propagator.}
\label{fig_vp13}
\end{figure}
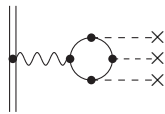

Other two vacuum polarization terms included in this work, namely the
K{\"a}ll{\'e}n and Sabry term \cite{206}, which contributes to the same order as
 the iterated Uelhing correction of Sec.~\ref{subsubsec:vpsc} and the Wichmann and Kroll term \cite{236}, were
calculated by perturbation theory. The numerical coefficients for both potentials are from \cite{277}. The Feynman diagrams of these terms
are shown, respectively, in Fig.~\ref{fig_vp21} and in
Fig.~\ref{fig_vp13}

%
%
\begin{figure}[htb]
\centering
\setlength{\unitlength}{0.27 mm}
\begin{picture}(260,60)(0,0)
\SetScale{.8}
%
\SetOffset(0,15)
  \Line(0,0)(0,50)
  \Line(3,0)(3,50)
  \Vertex(1.5,25){2}
  \Vertex(28.5,25){2}
  \Photon(3,25)(27,25){3}{3}
  \Vertex(38.5,15){2}
  \Vertex(38.5,35){2}
  \Photon(38.5,15)(38.5,35){2}{3}
  \CArc(38.5,25)(10,0,360)
  \Vertex(48.5,25){2}
  \DashLine(48.5,25)(68.5,25){3}
  \Text(74.5,26)[]{$\times$}
  \Text(100,26)[c]{+}
%
%
\SetOffset(120,15)
  \Line(0,0)(0,50)
  \Line(3,0)(3,50)
  \Vertex(1.5,25){2}
  \Vertex(28.5,25){2}
  \Photon(3,25)(27,25){3}{3}
  \CArc(38.5,25)(10,0,360)
  \Vertex(48.5,25){2}
  \Vertex(77,25){2}
  \Photon(48.5,25)(77,25){3}{3}
  \CArc(87,25)(10,0,360)
  \Vertex(97,25){2}
  \DashLine(97,25)(117,25){3}
  \Text(123,26)[]{$\times$}
%
\end{picture}
        \caption[]{Feynman diagrams for the two-loop vacuum
          polarization K{\"a}ll{\'e}n and Sabry contribution of the order of
          $\alpha^2(\alpha Z)$. The dashed lines starting with a
          $\times$ represent the interaction with the nucleus, the
          double line a bound kaon wave function or propagator and
          the wavy line a retarded photon propagator.}
\label{fig_vp21}
\end{figure}
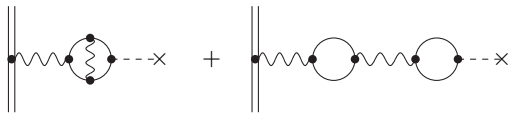 
%

%
%
%

%
%
\subsection{\label{subsec:eresult}Other corrections}

Other corrections contributes to the theoretical binding energy, $E$.
The energies obtained from the Klein-Gordon equation \eqref{eq:KGE} (with
finite nucleus and vacuum polarization correction) are already
corrected for the reduced mass $(1+m_{K-}/M_{A})$ where $m_{K-}$ is the Kaon mass and $M_{A}$ the \emph{total} mass of the system. Yet in the
relativistic formalism used here, there are other recoil to be
considered.  The first recoil correction is $-B^{2}/2M_{A}$, where $B$
is the binding energy of the level. For fermions other corrections are
known which are discussed, \emph{e.g.}, in \cite{765}. For bosons the
situation is not so clear. Contributions to the next order correction
can be found in \cite{aas83}.  Their expression depends on the nuclear
spin, and are derived only for spin $0$ and spin $1/2$ nuclei. For a
boson bound to a spin $0$ nucleus, this extra correction is non zero
except for $\ell=0$ states, and is of the same order as the
$-B^{2}/2M_{A}$ recoil term. For spin $1/2$ nuclei it is of order
$(Z\alpha)^ 4 m_{K-}^{2}/M_{A}^{2}$ and thus, being reduced by an
extra factor $m_{K-}/M_{A}$ should be negligible except for light
elements.  The self-energy for heavy particles is usually neglected.
To our knowledge no complete self-energy correction has been performed
for bosons. For deeply bound particles the vacuum polarization due to creation of virtual 
muon pairs become sizeable. We have evaluated it in the Uelhing approximation. This corrections is sizeable only for the deeply bound levels, and is always small 
compared to other corrections.

\subsection{\label{subsec:finpartsiz}Finite size of the bound particle}
Contrary to leptons like the electron or the muon, mesons like the pion or the kaon or baryons like the antiproton, are composite particles with non-zero charge distribution radii. These radii are of the same order of magnitude at the proton charge radius.
In order to take that correction into account we use a correction potential, that can be treated either as a perturbation or self-consistently. This correction is derived assuming that the nucleus and the particle are both uniformly charged spheres.
Denoting the nuclear charge radius by $R_{1}$, the particle charge radius by $R_{2}$ and the distance between the center of both charge distributions by $r$, one can easily derive (assuming $R_{1}\ge R_{2}$) 
\cite{bou98}:
\begin{equation}
        \left\{ 
        \begin{array}{ll}
            0 \leq r \leq (R_{1}-R_{2}): &
                V (r)= {\displaystyle - \frac{Z (-5 r^{2} + 15 
                R_{1}^{2} - 3 R_{2}^{2})}{10 R_{1}^{3}}} \\
                             (R_{1}-R_{2}) \leq r \leq (R_{1}+R_{2}):&
                 V (r)={\displaystyle \frac{Z}{160 r R_{1}^{3} 
                R_{2}^{3}} \left[ ( r^{6} - 15 r^{4} (R_{1}^{2}+R_{2}^{2}) + 
                40 r^{3} (R_{1}^{3}+ R_{2}^{3}) - 45 r^{2} 
                (R_{1}^{2}-R_{2}^{2})^{2}  \right.}\\
                &{\displaystyle \phantom{V(r)=)}\left. + 24 r 
                (R_{1}+R_{2})^{3} ( R_{1}^{2}-3R_{1} R_{2} + R_{2}^{2}) - 5 
                (R_{1}-R_{2})^{4} (R_{1}^{2}+4 R_{1} R_{2} +R_{2}^{2}) 
                \right]} \\
                (R_{1}+R_{2}) \leq r \leq \infty: &
                V(r) ={\displaystyle  - \frac{Z}{r}}
        \end{array} \right.
\end{equation}
For the Kaon we use a RMS radius of $0.560 \pm 0.031$~fm \cite{PDBook2004}. As an example we show in table \ref{tab:k_part}
 all the contributions included in the present work in the case of lead, including the kaon finite size and strong interaction shift from Ref. \cite{702}. 
 This correction is very large for deeply bound levels. Yet it remains small compared to hadronic corrections, which dominates heavily all other corrections except the Coulomb contribution.

\begin{table*} 
\squeezetable                     
\begin{center}                      
\caption{Contributions to the lower level energy of kaonic lead and comparison to hadronic shift and width \protect \cite{702} (MeV)). Experimental values
for transition energies are known for the $8k\to 7i$ transition and up (see Table~\protect\ref{tab:transition_energies})
 \label{tab:k_part}}
\scriptsize{
\begin{tabular}{cdddddddd}                              
\hline                              
\hline                              
& $1s$ & $2p$ & $3d$ & $4f$ & $5g$ & $6h$ & $7i$ & 8k \\
Coul. & -17.61031 & -12.70627 & -8.52297 & -5.43754 & -3.54036 & -2.45713 & -1.803475 & -1.379929 \\
Uehling ($ \alpha(Z \alpha) $)& -0.09971 & -0.08067 & -0.05924 & -0.03733 & -0.02091 & -0.01193 & -0.007218 & -0.004573 \\
Iter. Uehling ($ \alpha^{2}(Z \alpha) $) & -0.00017 & -0.00026 & -0.00030 & -0.00023 & -0.00012 & -0.00005 & -0.000027 & -0.000015 \\
Uehling (Muons)& -0.00001 & 0.00000 & 0.00000 & 0.00000 & 0.00000 & 0.00000 & 0.000000 & 0.000000 \\
Wich. \& Kroll ($ \alpha^{3}(Z \alpha) $) & 0.00181 & 0.00112 & 0.00079 & 0.00053 & 0.00034 & 0.00023 & 0.000157 & 0.000112 \\
K\"all\'en \& Sabry  ($ \alpha^{2}(Z \alpha) $) & -0.00085 & -0.00067 & -0.00048 & -0.00029 & -0.00016 & -0.00009 & -0.000051 & -0.000032 \\
Relat. Recoil & -0.00081 & -0.00042 & -0.00019 & -0.00008 & -0.00003 & -0.00002 & -0.000008 & -0.000005 \\
Part-size & 0.05435 & 0.03717 & 0.01600 & 0.00298 & 0.00017 & 0.00000 & 0.000000 & 0.000000 \\
Hadronic shift& 10.5 & 6.02 & 2.70 & 0.796 & 0.098 & 0.004 & & \\
Total & -7.2 & -6.73 & -5.87 & -4.676 & -3.463 & -2.465 & -1.810622 & -1.384443 \\
Hadronic Width & 1.42 & 1.38 & 0.848 & 0.468 & 0.118 & 0.009 & & \\
\hline                              
\hline                              
\end{tabular}
}
\end{center}
\end{table*}
%

%
%
\section{\label{sec:res_dis}Results and Discussion}

In Table \ref{tab:energy_levels} we compare the energy values
calculated in this work for selected kaonic atoms with existing
theoretical values.  It was assumed a kaon mass of $m_{K^{-}}$ =
493.677$\pm $0.013 MeV \cite{PDBook2004}. All energy values listed are in keV
units. The values obtained by other authors agree with ours to all
figures, even though earlier calculations are much less accurate.

In Table \ref{tab:transtion_energ_partial} we present the transition
energy values, obtained in this work, and by other authors, for the
Al($5g \rightarrow 4f$) and the Pb($12o \rightarrow 11n$)
transitionsn, in keV units. Again, we observe a good agreement between
all the listed results.

In Table \ref{tab:transition_energies} we list, in keV units, the
calculated kaonic atom X-ray energies for transitions between circular
levels $(n,\ell=n-1) \rightarrow (n-1,\ell=n-2)$ in this work. The
transitions are identified by the initial ($n_i$) and final ($n_f$)
principal quantum numbers of the pertinent atomic levels. The
calculated values of the transition energies in this work are compared
with available measured values ($E_m$), and with other calculated
values ($E_c$).



Comparison between the present theoretical values and the measured
values shows that the majority of our transition energies are inside
the experimental error bar. There are however a number of exceptions:
$2p \rightarrow 1s$ transition in hydrogen, the $3\to 2$ transition in
Be, the $4\to3$ transitions in Si and Cl, the $7\to 6$ transition in
tungsten and the U($8 \rightarrow 7$). In several cases the
measurement for transitions between the levels immediately above is in
good agreement.  These levels are less sensitive to strong
interaction, because of a smaller overlap of the kaon wave function
with the nucleus.  This thus point to a strong interaction effect.
This is certainly true for hydrogen in which the $1s$ state is
involved, and quite clear for Be.  In the case of lead, the large
number of measured transitions makes it interesting to look into more
details, and investigate the eventual role of electrons.

To assess the influence of the electrons that survived to the cascade
process of the kaon, we calculated transition energies in two cases
for which experimental measurements have small uncertainties, and a
long series of measured transitions, namely the Pb ($12o \rightarrow
11n$) and ($13q \rightarrow 12o$), without and with electrons. In
Table \ref{tab:transtions_pb} we list, in units of eV,
the transition energy contributions due to the inclusion of 1, 2, 4
and 10 electrons in the kaonic system, for the mentioned transitions.
%
%
We conclude that the electron screening effect by $1s$ electrons is
much larger than experimental uncertainties, while the effect of $2s$
electrons is of the same order.  Other electrons have a negligible
influence.  Moreover, we can conclude that the electronic correlation
effects are negligible since the transition energy contribution of the
Be-like system for configuration $1s^2 2s^2$ differs only by 0.1~eV
from the energy obtained with the $1s^2 2s^2 + 1s^2 2p^2$
configurations, which represents the well-known strong intrashell
correlation of Be-like ions.

%
We investigated the electronic influence in few more transitions,
using the same guideline to choose the more relevant ones, i.e., small
experimental uncertainties.
We present in Table \ref{tab:transition_energies_electrons} the
differences between the measured, $E_{\mbox{m}}$, and the calculated
transition energies, $E$, without electrons and with 1, 2, 3, 4 and 18
electrons, in units of eV.  The transitions between ($n$, $\ell =n-1$)
states are identified by the initial ($n_i$) and final ($n_f$)
principal quantum numbers of the atomic levels. In the calculated
energies for lead, we included the $\alpha(Z\alpha)^{5,7}$ vacuum
polarization and nuclear polarization from Ref.~\cite{702}. To our
knowledge, this information is not available for other nuclei. When
there are more than one measurement available, we use the weighted
average for both the transition energy $E_{\mbox{m}}$ and the
associated uncertainty $\Delta E_{\mbox{m}}$, assuming normally
distributed errors. The calculated values are compared with the
experimental uncertainty values $\Delta E_m$.
For the Pb transitions we can use the results of Table
\ref{tab:transition_energies_electrons} to estimate the number of
residual electrons for different transitions.  For the $9\to 8$
transitions, our results shows that there must remain at least two
electrons, and are compatible with up to 18 remaining electrons. This
is more or less true for all the other transitions except for the
$8\to 7$.


%

\begin{figure}[htb]
\includegraphics[height=6cm, width=8.5cm]{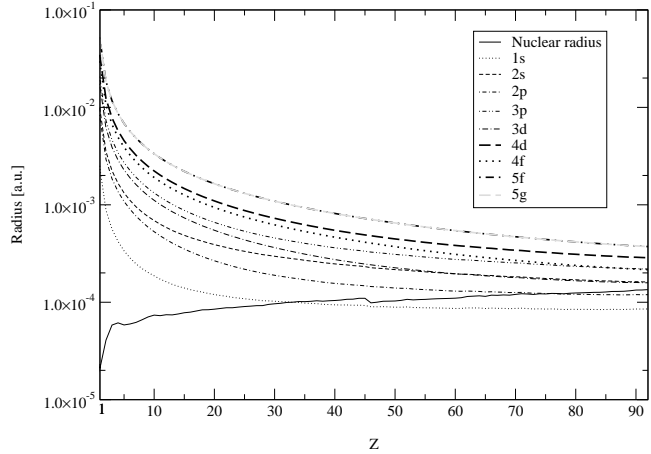}
\caption{Nuclear radius and average radius of the kaonic atoms levels
for $Z=1-92$. 
\label{figure_3}}
\end{figure}

In Figure \ref{figure_3} we plot the nuclear radius and the average
radius of some kaonic atoms wavefunction as function of $Z$. This
graph shows when the wavefunction radius and the nuclear radius are of
the same order of magnitude.  It can be used to find which levels are
most affected by the strong-interaction for a given $Z$ value.
%
%





%
%

\section{\label{sec:conclusion}Conclusion}

In this work we have evaluated the energies of the circular $(n,\ell
=n-1)$, $1 \le n \le 12$ and first parallel $(n,\ell =n-2)$ levels for
several (hydrogenlike) kaonic atoms throughout the periodic
table. These energy levels were used to obtain transition energies to
compare to the available experimental and theoretical cases.

Our transition energy calculations reproduce experiments on kaonic
atoms within the error bar in the majority of the cases. In all the
cases the theoretical values are more accurate than experimental ones,
as the experiments face the X-ray contamination by other elements.

We have investigated the overlap of the nuclear radius and average
radius of kaonic levels as function of $Z$ and the influence of
electrons that survived the cascade process on the transition
energies.

%
%
\begin{acknowledgments}
  This research was supported in part by FCT project POCTI/FAT
  /44279/2002 financed by the European Community Fund FEDER.
  Laboratoire Kastler Brossel is Unit\' e Mixte de Recherche du CNRS
  n$^\circ$ C8552.
\end{acknowledgments}
%
%
%
\bibliographystyle{apsrev}
\bibliography{jps}

%

%
                            

{\footnotesize    

\begin{table*}    
\begin{center}    
%
%
\caption{  Calculated binding energies of various atomic levels for several kaonic atoms in keV units.  
\protect\label{tab:energy_levels}}    
%
\begin{tabular}{ccrrr}                                    
%
%
\hline                                    
\hline                                    
\\                                    
Nucleus    &  Atomic level  &  This work     &  Ref. \cite{685}     &  Ref. \cite{675}     \\       
\\                                    
\hline                                    
\\      
H & $1s$ & 8.63360 & 8.634 & $ $ \\
& $2p$ & 2.15400 & 2.154 & $ $ \\
& $3p$ & 0.95720 & 0.957 & $ $ \\
C & $1s$ & 430.815 & & $ 432 \times 10^0 $ \\
& $2s$ & 110.579 & & $ 111 \times 10^0 $ \\
& $2p$ & 113.762 & & $ 114 \times 10^0 $ \\
Ca & $1s$ & 3368.58 & & $ 344 \times 10^1 $ \\
& $2s$ & 1042.89 & & $ 106 \times 10^1 $ \\
& $2p$ & 1287.20 & & $ 129 \times 10^1 $ \\
& $3p$ & 573.356 & & $ 57 \times 10^1 $ \\
& $3d$ & 579.932 & & $ 58 \times 10^1 $ \\
& $4d$ & 325.871 & & $ 32 \times 10^1 $ \\
Pb & $1s$ & 17655.61 & & $ 1770 \times 10^1 $ \\
& $2s$ & 9372.797 & & $ 939 \times 10^1 $ \\
& $2p$ & 12749.89 & & $ 1277 \times 10^1 $ \\
& $3p$ & 6830.710 & & $ 684 \times 10^1 $ \\
& $3d$ & 8566.297 & & $ 857 \times 10^1 $ \\
& $4d$ & 4938.662 & & $ 494 \times 10^1 $ \\
& $4f$ & 5471.923 & & $ 548 \times 10^1 $ \\
& $5f$ & 3493.043 & & $ 349 \times 10^1 $ \\
& $5g$ & 3561.058 & & $ 356 \times 10^1 $ \\
& $6g$ & 2471.161 & & $ 247 \times 10^1 $ \\
& $6h$ & 2468.987 & & $ 247 \times 10^1 $ \\
& $7h$ & 1813.611 & & $ 182 \times 10^1 $ \\\\                                    
\hline                                    
\hline                                    
\end{tabular}                                    
\end{center}                                    
\end{table*}                                    
}                                    
%

\begin{table*}                                
{\footnotesize                                
\begin{center}                                
%
%
\caption{Calculated transition energies and respective contributions,
  in keV units, for the Al ($5g \rightarrow 4f$) and the Pb ($12o
  \rightarrow 11n$) transitions. The Al and Pb measured values
  ($E_{\mbox{m}}$) are from Ref.~\cite{704} and from Ref.~\cite{702},
  respectively.
%
\protect\label{tab:transtion_energ_partial}}                                        
%
\begin{tabular}{ldrrdrr}                                        
%
%
\hline                                        
\hline                                        
\\                                        
      &  \multicolumn{2}{c}{Al ($5g \rightarrow 4f$)}       & &  \multicolumn{3}{c}{Pb ($12o \rightarrow 11n$)}        \\              
        \cline{2-3}          \cline{5-7}        \\              
      &  \multicolumn{1}{c}{This work}  & Ref. \cite{701}    & &  \multicolumn{1}{c}{This work}  & Ref. \cite{702} & Ref. \cite{707}   \\                
\\   
\hline  \\                
\\     
 Coulomb     & 49.03755  & 49.04    & &  116.5666  & 116.575 & 116.600   \\                
 Vacuum Polarization     &    &     & &    &  &    \\                
 \hspace*{0.5cm}  $ \alpha(Z \alpha) $ &  0.19048  &     & &  0.4203  & 0.421 &    \\                
 \hspace*{0.5cm}  $ \alpha(Z \alpha)^3 $ &  -0.00012  &     & &  -0.0109  & -0.011 &    \\                
 \hspace*{0.5cm}  $ \alpha^2(Z \alpha) $ &  0.00181  &     & &  0.0040  & 0.003 &    \\                
 \hspace*{0.5cm}   Others  &    &     & &    & -0.002 &    \\                
 \hspace*{0.5cm}   Total  &  0.19218  & 0.19    & &  0.4135  & 0.412 & 0.410   \\                
 Recoil     &  0.00022  &     & &  0.0004  &  &    \\                
 Others     &    &     & &    & -0.044 & -0.050   \\                
 Total     &  49.22994  & 49.23    & &  116.9804  & 116.943 & 116.960   \\                
\\
\hline  \\                
 $ E_{\mbox{m}}$        &  49.249 (19) & & & 116.952 (10) \\
\\                                          
\hline                                          
\hline                                      
\end{tabular}                                      
\end{center}                                      
}                                      
\end{table*}                                      
%
                                      

%
\begin{table*}                                                                                                               
{\footnotesize  

\begin{center} 
                                 
%
\caption{Kaonic atom X-ray energies of circular transitions $(n,\ell=n-1)
  \rightarrow (n-1,\ell=n-2)$ in keV units. The transitions are
  identified by the initial ($n_i$) and final ($n_f$) principal
  quantum numbers of the atomic levels. The calculated values of the
  transition energies in this work are compared with the available
  measured values ($E_{\mbox{m}}$) and with other calculated values
  ($E_{\mbox{c}}$).
\protect\label{tab:transition_energies}}                                        
%
\begin{tabular}{ccrdrc}                                           
%
%
\hline                                           
\hline                                           
\\                                           
Nucleus    &   Transition   &  This work    &  E_{\mbox{m}}  & $E_{\mbox{c}}$ & Ref.   \\        
    &   $n_i \rightarrow n_f$   &      &        &       &    \\        
\\                                           
\hline                                           
\\                                           
$^{  }$ H  & $ 2 \rightarrow 1 $ &  6.480    &  6.675    \ ( 60 )  &  $ 6.482    $ & \cite{704}   \\        
$^{  }$    & $    $              &           &  6.96     \ ( 9 )   &  $     $ & \cite{753}   \\      
$^{  }$ He & $ 3 \rightarrow 2 $ &  6.463    &  6.47     \ ( 5 )   &  $ 6.47         $ & \cite{769}   \\      
$^{  }$    & $ 4 \rightarrow 2 $ &  8.722    &  8.65     \ ( 5 )   &  $ 8.73         $ & \cite{769}   \\      
$^{  }$ Li & $ 3 \rightarrow 2 $ &  15.330   &  15.320   \ ( 24 )  &  $ 15.319    $ & \cite{703}   \\      
$^{  }$    & $    $              &           &  15.00    \ ( 30 )  &   15.28    & \cite{706}   \\      
$^{  }$    & $ 4 \rightarrow 2 $ &  20.683   &  20.80    \ ( 30 )  &  $ 20.63    $ & \cite{706}   \\      
$^{  }$ Be & $ 3 \rightarrow 2 $ &  27.709   &  27.632   \ ( 18 )  &  $ 27.632    $ & \cite{703}   \\      
$^{  }$    & $    $              &           &  27.50    \ ( 30 )  &   27.61     & \cite{706}   \\      
$^{  }$    & $ 4 \rightarrow 3 $ &  9.677    &  9.678    \ ( 1 )   &  $ 9.678    $ & \cite{704}   \\      
$^{  }$ C  & $ 4 \rightarrow 3 $ & 22.105    &  22.30    \ ( 30 )  &  $ 22.06    $ & \cite{706}   \\      
$^{  }$ Al & $ 4 \rightarrow 3 $ & 106.571   &  106.45   \ ( 5 )   &  $ 106.58    $ & \cite{701}   \\      
$^{  }$    & $ 5 \rightarrow 4 $ & 49.230    &  49.27    \ ( 7 )   &  $ 49.23    $ & \cite{701}   \\      
$^{  }$    & $    $              &           &  49.249   \ ( 19 )  &   49.233     & \cite{704}   \\      
$^{  }$    & $ 6 \rightarrow 5 $ & 26.707    &  26.636   \ ( 28 )  &  $ 26.685    $ & \cite{704}   \\      
$^{  }$    & $ 9 \rightarrow 8 $ & 7.151     &  7.150    \ ( 1 )   &  $ 7.150    $ & \cite{704}   \\      
$^{  }$ Si & $ 4 \rightarrow 3 $ & 123.724   &  123.51   \ ( 5 )   &  $ 123.75    $ & \cite{701}   \\      
$^{  }$    & $ 5 \rightarrow 4 $ & 57.150    &  57.23    \ ( 7 )   &  $ 57.16    $ & \cite{701}   \\      
$^{  }$ Cl & $ 4 \rightarrow 3 $ & 183.287   &  182.41   \ ( 40 )  &  $ 183.35    $ & \cite{707}   \\      
$^{  }$    & $ 5 \rightarrow 4 $ & 84.648    &  84.44    \ ( 26 )  &  $ 84.67    $ & \cite{707}   \\      
$^{  }$ K  & $ 5 \rightarrow 4 $ & 105.952   &  105.86   \ ( 28 )  &  $ 105.97    $ & \cite{707}   \\      
$^{  }$ Ca & $ 5 \rightarrow 4 $ & 117.466   &  117.64   \ ( 22 )  &  $ 117.48    $ & \cite{707}   \\      
$^{  }$ Ti & $ 5 \rightarrow 4 $ & 142.513   &  141.8            & &\cite{771}   \\      
$^{  }$ Ni & $ 5 \rightarrow 4 $ & 231.613   &  231.49   \ ( 7 )   &  $ 231.67    $ & \cite{701}   \\      
$^{  }$    & $ 6 \rightarrow 5 $ & 125.563   &  125.60   \ ( 5 )   &  $ 125.59    $ & \cite{701}   \\      
$^{  }$    & $ 7 \rightarrow 6 $ & 75.606    &  75.59    \ ( 5 )   &  $ 75.62    $ & \cite{701}   \\      
$^{  }$ Cu & $ 5 \rightarrow 4 $ & 248.690   &  248.50   \ ( 22 )  &  $ 248.74    $ & \cite{701}   \\      
$^{  }$    & $ 6 \rightarrow 5 $ & 134.814   &  134.84   \ ( 5 )   &  $ 134.84    $ & \cite{701}   \\      
$^{  }$    & $ 7 \rightarrow 6 $ & 81.173    &  81.15    \ ( 5 )   &  $ 81.18    $ & \cite{701}   \\      
$^{  }$ Mo & $ 8 \rightarrow 7 $ & 110.902   &  110.90   \ ( 28 )  &  $ 110.92    $ & \cite{707}   \\        
$^{  }$ Rh & $ 8 \rightarrow 7 $ & 127.392   &  127.43   \ ( 31 )  &  $ 127.17    $ & \cite{707}   \\        
$^{  }$    & $ 9 \rightarrow 8 $ & 87.245    &  87.25    \ ( 35 )  &  $ 86.66    $ & \cite{707}   \\
$^{  }$ Nd & $ 9 \rightarrow 8 $ & 155.569   &  155.60   \ ( 29 )  &  $ 155.63    $ & \cite{707}   \\
$^{  }$    & $10 \rightarrow 9 $ & 111.159   &  110.84   \ ( 32 )  &  $ 111.17    $ & \cite{707}   \\
$^{  }$ W  & $ 7 \rightarrow 6 $ & 535.240   &  534.886  \ ( 92 )  &  $ 535.239    $ & \cite{754}   \\
$^{  }$    & $ 8 \rightarrow 7 $ & 346.571   &  346.624  \ ( 25 )  &  $ 346.545    $ & \cite{754}   \\
$^{  }$ Pb & $ 8 \rightarrow 7 $ & 426.180   &  426.181  \ ( 12  ) &  $ 426.201    $ & \cite{702}   \\
$^{  }$    & $    $              &           &  426.221  \ ( 57 )  &   426.149     & \cite{754}   \\
$^{  }$    & $ 9 \rightarrow 8 $ &  291.626  &  291.577  \ ( 13 )  &  $ 291.621    $ & \cite{702}   \\
$^{  }$    & $    $              &           &  291.74   \ ( 21  ) &   291.59     & \cite{707}   \\
$^{  }$    & $ 10 \rightarrow 9$ &  208.298  &  208.256  \ ( 8 )   &  $ 208.280    $ & \cite{702}   \\
$^{  }$    & $    $              &           &  208.69   \ ( 21  ) &  $ 208.34    $ & \cite{707}   \\
$^{  }$    & $11 \rightarrow 10$ &  153.944  &  153.892  \ ( 11 )  &  $ 153.916    $ & \cite{702}   \\
$^{  }$    & $    $              &           &  154.13   \ ( 21 )  &  $ 153.94    $ & \cite{707}   \\
$^{  }$    & $12 \rightarrow 11$ &  116.980  &  116.952  \ ( 10 )  &  $ 116.943    $ & \cite{702}   \\
$^{  }$    & $    $              &           &  116.96   \ ( 25 )  &  $ 116.96    $ & \cite{707}   \\
$^{  }$    & $13 \rightarrow 12$ &  90.970   &  90.929   \ ( 15 )  &  $ 90.924    $ & \cite{702}   \\
$^{  }$ U  & $ 8 \rightarrow 7 $ &  537.442  &  538.315  \ ( 100 ) &  $ 538.719    $ & \cite{754}   \\      
\\                                         
\hline                                         
\hline                                         
\end{tabular}                                         
\end{center}                                         
}                                         
\end{table*}                                         
%
                                         

%
\begin{table*}                          
\begin{center}               
\caption{Transition energy contributions due to the inclusion of 1, 2,
  4 and 10 electrons, in eV units, for Pb ($12o \rightarrow 11n$) and
  ($13q \rightarrow 12o$) transitions.  Effect of the intra-shell
  correlation $1s^2 2s^2+ 1s^2 2p^2$ in the Be-like case is presented in the column labeled [Be] corr.
\protect\label{tab:transtions_pb}}                
\begin{tabular}{lrrrrr}                
\hline                
\hline                
\\                
&\multicolumn{5}{c}{Transition Energy Contributions}   \\
\cline{2-6}  \\                                     
& [H] & [He] & [Be] & [Be] corr.& [Ne]  \\
\\
\hline  \\
 \\
$12o \rightarrow 11n$ & -20.735 & -40.917 & -47.419 & -47.334 & -47.484  \\
$13q \rightarrow 12o$ & -24.126 & -47.609 & -55.135 & -55.041 & -55.245  \\
\\                
\hline                
\hline                
\end{tabular}                
\end{center}                
%
\end{table*}                
%

\begin{table*}                                                         
%
\begin{center}                               
\caption{Differences between the measured, $E_{\mbox{m}}$, and the
  calculated transition energies values, $E$, without electrons and
  with 1, 2, 3, 4 and 18 electrons, respectively, in eV units.  The
  transitions between ($n$, $\ell =n-1$) states are identified by the
  initial ($n_i$) and final ($n_f$) principal quantum numbers of the
  atomic levels. $\Delta E_{\mbox{m}}$ stands for the experimental
  uncertainty. For the cases in which there are more than one measure,
  the weighted average was taken for both $E_{\mbox{m}}$ and $\Delta
  E_{\mbox{m}}$. In the case of the [Li] core the calculation was made for $J=\ell_{\mbox{kaon}}+1/2$.
\protect\label{tab:transition_energies_electrons}}                                       
\begin{tabular}{ccrrrrrrrrc}                                       
%
%
\hline                                       
\hline                                       
\\                                       
    &   Transition   &     &     \\                                           
Nucleus & $n_ i\rightarrow n_f$ & \multicolumn{6}{c}{$E_{\mbox{m}}-E$} & $ \Delta E_{\mbox{m}}  $ & Refs.   \\
\cline{3-8}  \\                                     
   &     &  No Elect.  &  [H]  &  [He]  & [Li] & [Be] & [Ar] $   $ &    \\                                         
\\                                     
\hline                                     
\\                                     
 Be & $ 4 \rightarrow 3 $ & $ 0.5 $ & $ 0.5 $ & $ 0.5 $ & 0.5 &&&$ 1  $ & \cite{703} \\      
 Al & $ 9 \rightarrow 8 $ & $ -1.0 $ & $ -0.4 $ & $ 0 $ & 0.1 &  && $ 1  $ & \cite{701} \\      
 Pb & $ 8 \rightarrow 7 $ & $ -9 $ &  $ 0   $ & $ 8   $ & $  10$ &$11$&$12$& $ 12  $ & \cite{754,702} \\      
    & $ 9 \rightarrow 8 $ &   $ -49 $ & $ -37 $ & $ -26 $ & $ -25$ &$-23$&$-22$& $ 13  $ & \cite{707,702} \\      
    & $ 10 \rightarrow 9 $ &  $ -41 $ & $ -26 $ & $ -12 $ & $ -10$ &$-8$&$-7$& $ 8  $ & \cite{707,702} \\      
    & $ 11 \rightarrow 10 $ & $ -50 $ & $ -33 $ & $ -16 $ & $ -13$ &$-10$&$-9$& $ 11  $ & \cite{707,702} \\      
    & $ 12 \rightarrow 11 $ & $ -27 $ & $ -6  $ & $ 14  $ & $ 17 $ &$20$&$22$& $ 10  $ & \cite{707,702} \\      
    & $ 13 \rightarrow 12 $ & $ -40 $ & $ -10 $ & $ 8   $ & $ 12 $ &$15$&$17$& $ 15  $ & \cite{707,702} \\      
\\                                     
\hline                                     
\hline                                     
\end{tabular}                                     
\end{center}                                     
%
\end{table*}                                     
%

%
%
%

\end{document}